\documentclass[aip,twocolumn,letterpaper]{revtex4}

\pdfoutput=1  
\usepackage[utf8x]{inputenc}  

\usepackage{fullpage}
\usepackage{xcolor}

\usepackage{graphics}
\usepackage{epsfig}

\usepackage[margin=1.0in]{geometry}
\usepackage{url}

\begin{document}
\title{Carbon Dioxide, Fusion, and Stellarators}
\author{Allen H Boozer}
\affiliation{Columbia University, New York, NY  10027\\ ahb17@columbia.edu}

\begin{abstract}
Much emotion is expended on the dangers of carbon dioxide, but solutions require reason and recognition of facts: (1) The cost of developing options is approximately a thousand times less than their deployment.  (2) Timescales involve two questions: (a) How quickly can an option be demonstrated? (b) How quickly can the required equivalent of thousands of units be built. Two questions are implied: (1) What options would most fundamentally change the carbon-dioxide problem? (2) For each option, how could it be demonstrated most quickly?  An option of fundamental importance is direct air capture of carbon dioxide. The option that appears most attractive for carbon-free energy production is the stellarator fusion concept, which is poised for a rapid demonstration.

\end{abstract}

\date{\today} 
\maketitle

\section{Introduction}

\begin{figure}
\centerline{ \includegraphics[width=3.0in]{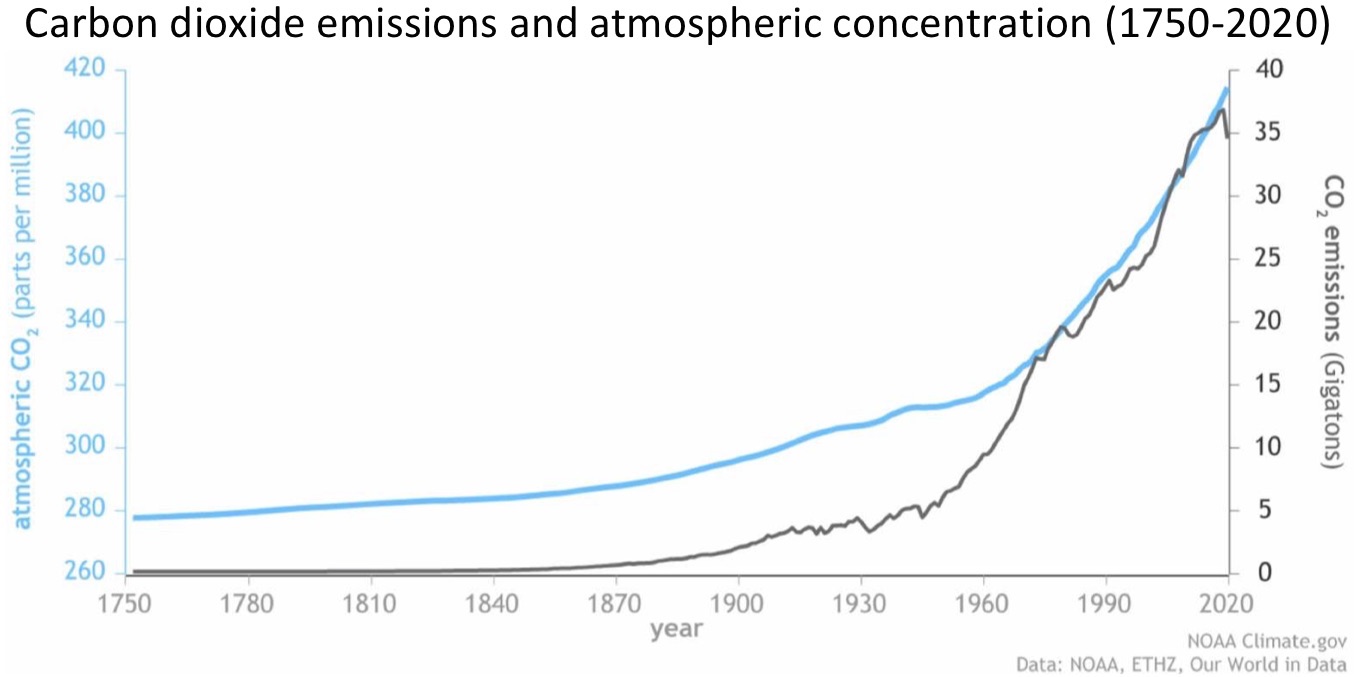} }
\caption{The rate of CO$_2$ emissions is doubling approximately every thirty years and the enhancement in the atmospheric concentration above its pre-industrial level is doubling approximately every forty years.   This NOAA Climate.gov graph by Rebecca Lindsey \protect\cite{CO2} was adapted from the original by Howard Diamond (NOAA ARL). Atmospheric CO$_2$ data from NOAA and ETHZ. CO$_2$ emissions data from Our World in Data and the Global Carbon Project. }
\label{fig:CO2}
\end{figure}

The importance of fast and high-certainty development of fusion energy is defined by the exponential increase in both carbon dioxide emissions and the enhancement of the atmospheric concentration, Figure \ref{fig:CO2}.  When the speed and certainty of development are the primary criteria, the stellarator is the choice of fusion concept \cite{CO2-Stell,Stell-path}.     

The increased CO$_2$ concentration is associated with an increasing atmospheric temperature, ocean level, and ocean acidity.  Uncertainties in our understanding raise additional concerns.  

Figure \ref{fig:CO2} implies action must be taken, and 2050 is generally taken as the date by which net CO$_2$ emissions must end  \cite{IRENA:2021,WEO:2021}.  The focus has not been on controlling atmospheric CO$_2$  but on eliminating fossil fuels (coal, oil, and natural gas).

What is not recognized is the enormous ratio between the cost of deployment of an option for a solution to the cost of development of the option to the point of deployment.  A typical ratio is the one for fusion energy, approximately a thousand.  Only one demonstration fusion power plant is required to determine the properties of fusion power, but approximately ten thousand fusion power plants are required to affect world energy production.  

The cost of deployment using existing options to achieve net-zero emissions by 2050 is high.  Page 47 of a 2021 International Energy Agency report \cite{WEO:2021} says four trillion dollars per year will be required by 2030.  A hundred times smaller expenditure, forty billion dollars a year, could provide better options through well-organized development programs.  By comparison, the size of the world energy industry is approximately six trillion dollars a year.  

\begin{figure}
\centerline{ \includegraphics[width=2.2in]{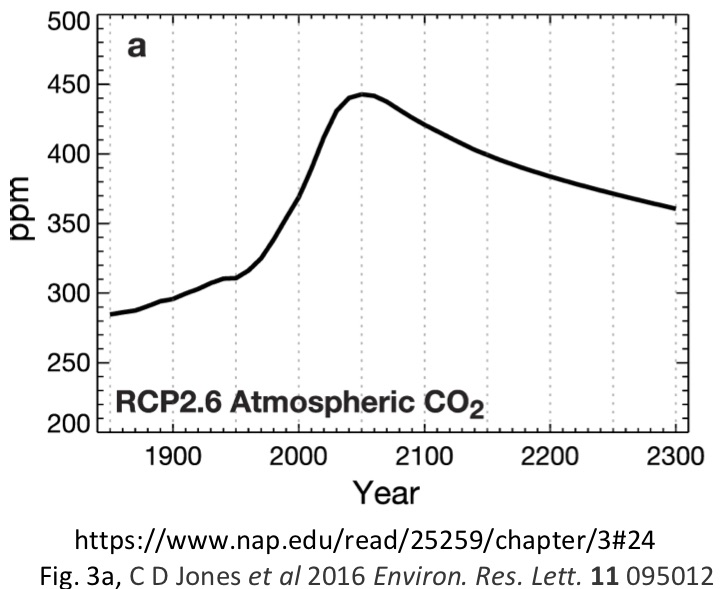} }
\caption{An enhanced carbon dioxide concentration only slowly decreases towards its natural level, see page 24 of the U.S. National Academy Report  \protect\cite{NAS-CO2}, which cites \protect\cite{Jones:2016}.}
\label{fig:CO2 decay}
\end{figure}

Controlling emissions alone is expensive but leaves the CO$_2$ concentration above its present level for centuries, Figure \ref{fig:CO2 decay}.  The U.S. National Academy and the British Royal Society state on page 22 of their 2020 report \cite{Longterm effects}: ``\emph{Even if emissions of greenhouse gases were to suddenly stop, Earth's surface temperature would require thousands of years to cool and return to the level in the pre-industrial era.}" These are timescales for climate-change  accommodation, not avoidance.

On 5 November 2021, U.S. Secretary of Energy, Jennifer Granholm announced \cite{DAC} the goal of building a direct air capture and sequestration system (DAC) at a gigaton level by 2050 with a CO$_2$ removal cost of less than \$100/ton.  The envisioned cost of removal and sequestration is standardly given \cite{NAS-CO2,Gertner:2021} as \$100 to \$200 per ton of CO$_2$, and research is required to achieve costs even in this range.  

To have a profound effect on arresting the level of CO$_2$ or the timescale for lowering that level requires a direct air capture system comparable in scale to the thirty-six gigatons that are now being emitted each year.  This is also the maximum emission rate under the different CO$_2$ mitigation plans described on page 33 of \cite{WEO:2021}.  The removal of thirty-six gigatons per year is far beyond what can be accomplished by measures such as planting trees \cite{NAS-CO2}.

If a one gigaton DAC system can be built by 2050, it would only be a matter of will whether a thirty-six gigaton system could be completed shortly thereafter.  At \$100/ton, present emissions could be removed for less than the four trillion dollars a year required to end emissions by 2050 as envisioned by the International Energy Agency \cite{WEO:2021}.

The effect on the timescale for maintaining or returning to any earlier atmospheric concentration of carbon dioxide using direct air capture is profound.  When it is assumed CO$_2$ emissions are ended by mid-century, the sum of all CO$_2$ emissions due to humans is approximately fifty times larger than the thirty-six gigatons that are now being emitted per year.  A CO$_2$ removal system that can remove as much carbon dioxide in a year as the highest one-year emission would eliminate any further increases in the level of atmospheric CO$_2$ once that system is available and reduce the enhancement of the CO$_2$ concentration on the timescale of a lifetime, not millennia.  


The time to develop a new option need not be too long to be consistent with the 2050 date for the implementation of a solution.  It took only fifteen years to go from the splitting of uranium in a laboratory to fission powered submarines.  Will and organization are critical.  Fears associated with World War II and the Cold War provided the necessary will.  General Groves and Admiral Rickover provided required organizational skills.

Sufficient will exists for large expenditures to address the CO$_2$ problem.  In 2020, one country, Germany, spent thirty-eight billion dollars subsidizing green energy \cite{German-subsidy}.  Organization is more difficult.  Without appropriate organization, a research program can expend arbitrarily large resources and take an arbitrarily long time. 

Since the cost of development is trivial compared to the cost of deployment, a rational world would ask what options would allow carbon-dioxide issues to be addressed with the greatest certainty, on the shortest possible timescale, and with the least detriment to the world economy.  Two such options are the direct removal of carbon dioxide \cite{NAS-CO2,Gertner:2021} and fusion \cite{CO2-Stell,Stell-path}.  Without clairvoyance, an optimal program must explore options that are never deployed. 

Page 4 of the 2019 U.S. National Academy report \cite{NAS-CO2} mentioned avoiding the moral hazard of ``\emph{reducing humanity's will to cut emissions in the near term}" by proposals for research on attractive options.   Discouraging the development of better options not only seems irresponsible but would likely delay the restoration of a desirable CO$_2$ level.  Reason is a better guide than emotion in determining when and how the switch from the development to the deployment of the best options should occur.  

What is meant by a desirable carbon-dioxide level is subtle; each level has winners and losers.  For example, what level is optimal for worldwide food production versus the flooding of low-lying regions.  In any case, global warming sounds far less dangerous than global cooling---a new ice age.  The last ice age ended approximately twelve thousand years ago.  People have inhabited the Earth for more than twenty times longer, and the Earth itself is three million times older.  

The optimal CO$_2$ level is ultimately a political question.  The optimum is often assumed to be the pre-industrial level.  Based on the primary planning documents, a return to the pre-industrial CO$_2$ concentration would take miillinnia \cite{Longterm effects}.  That is why the announcement by Secretary Granholm \cite{DAC} on direct air removal of CO$_2$ is of such great potential importance.

To be widely accepted, energy sources must be reliable and consistent with an increasing worldwide standard of living.  Eliminating the use of fossil fuels before acceptable alternatives are available is both expensive and counterproductive.  The February 2021 collapse in the electricity grid in Texas and the late-summer 2021 lull in the North Sea winds  \cite{Wind stopped} illustrate problems that occur when insufficient thought goes into ensuring system stability and providing backups for intermittent energy sources.   

The cost of energy is important, but longterm reliability is even more so.  Being without electricity for home lighting and heating during a few randomly occurring weeks a year is unacceptable.  The higher cost of a reliable energy source may be offset by efficiency measures.   Unreliability is addressed by home generators, but they are polluting, inefficient, and inequitably available.   

Wind and solar are the widely acclaimed alternative to fossil fuels.  Their practicality is location dependent; they are not universally applicable without long-distance transmission.  Their intermittency necessitates backup systems.  Batteries can be used for hours-long interruptions, but longterm interruptions, such as the lull in the North Sea winds, requires an alternative power source.  Natural gas turbines are the basis of the only system that is inexpensive, has quick turn-on and turn-off timescales, and is not location dependent.  Nevertheless, even natural gas systems require careful design for stability as illustrated by the Texas blackout.   Natural gas could be replaced by a manufactured carbon-free product, such as hydrogen.  If wind and solar were as reliable as often implied \cite{Burden} the use of the backup would be rare.  In addition, the carbon dioxide could be removed from the exhaust.

The only energy system that can be employed at the required scale while not being intermittent and location dependent is nuclear energy.  Nuclear energy, whether fission or fusion, has a low fuel but high capital cost.  Once built, the power plant should be operated as close to full power as large a fraction of the time as practical to recover the expense of construction.  Consequently, nuclear energy is not a practical backup for wind and solar.  Fission energy has waste, safety, and proliferation issues, which can be largely avoided with fusion,  but fusion has not yet been demonstrated.  

Carbon-dioxide control defines the need to develop fusion energy with the highest certainty and in minimum time.  Section \ref{sec:fusion} outlines how this can be done and why the stellarator, not the tokamak, provides the obvious path.


\begin{figure*}
\centerline{ \includegraphics[width=7in]{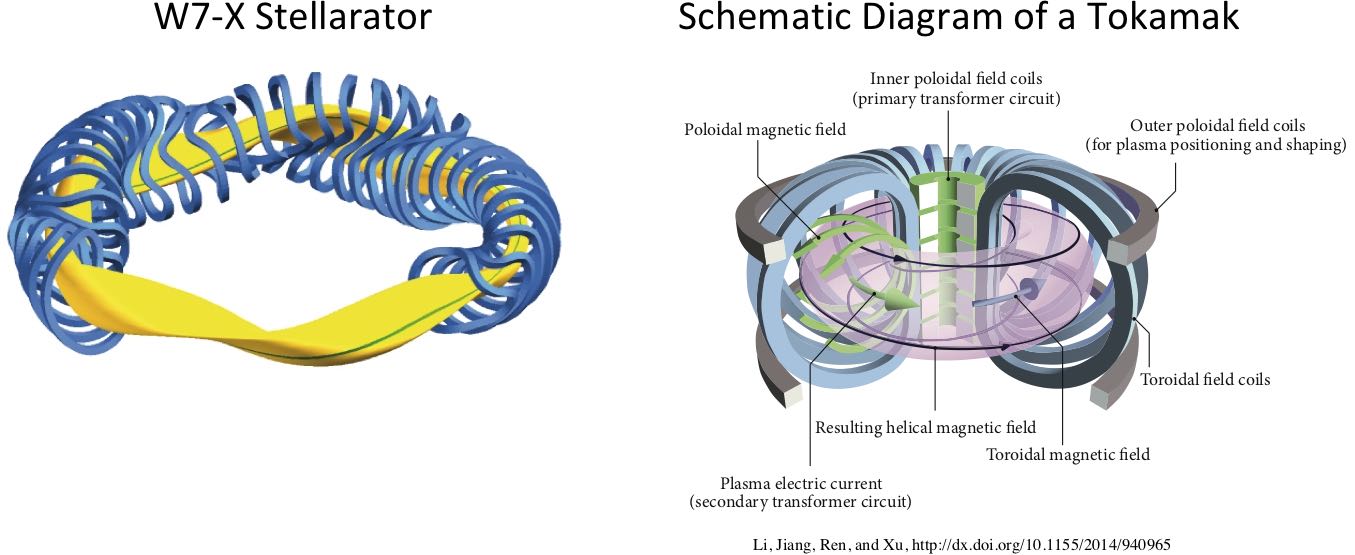} }
\caption{Both stellarators and tokamaks are toroidal, but stellarators have a helical twist and tokamaks are ideally axisymmetric.  The blue coils give the magnetic field that confines the yellow plasma in the stellarator diagram.  The magnetic field that confines the purple plasma in the tokamak diagram \protect\cite{Jiang} requires a current that flows toroidally in the plasma.}
\label{fig:stellarators-tokamaks}
\end{figure*}

\section{Fusion Energy \label{sec:fusion}}

Two types, or isotopes, of hydrogen, deuterium (D) and tritium (T), will react, called burning, to produce an ordinary helium and a neutron as well as a large amount of energy.  For this to happen, the temperature must be approximately a hundred million degrees centigrade, which in the conventional units of plasma physics is ten thousand electron volts, 10~keV.  At this temperature the electrons and ions separate to form a plasma, which is an ideal gas that is an excellent conductor of electricity.  The number density of the electrons and the ions is approximately $10^{20}$~m$^{-3}$, which implies the plasma has a pressure of approximately three atmospheres.   Each electron and ion moves approximately 10~km between interactions with other particles that change its momentum and energy.  These interactions are called collisions.  The plasmas in fusion power plants have scales of a few meters; the motion of particles on this scale is determined by the classical mechanics of collisionless particles in large-scale electric and magnetic fields.  For energy release from the DT to be adequate to maintain a DT burn in a power plant, the confinement time of energy in the plasma must be orders of magnitude longer than the timescale for collisions.  The implication is that the electron and ion velocities will be in the Maxwellian distribution that is characteristic of an ideal gas.  Tritium does not naturally occur in nature but can be produced by the neutron reacting with lithium in a blanket that surrounds the plasma.

The plasmas in stellarators and tokamaks are toroidal, Figure \ref{fig:stellarators-tokamaks}.  Magnetic field lines lie on nested toroidal surfaces.  The only way the plasma can escape is to drift or diffuse across
these surfaces.   The magnetic surfaces in stellarators can be defined by currents in coils that lie outside the plasma, Figure \ref{fig:stellarators-tokamaks}, but in tokamaks a current within the plasma has an essential role in formation of these surfaces.  Tokamak plasmas can in principle be exactly axisymmetric; stellarators must have helical shaping.

Confinement of the magnetic field lines and the particles on toroidal surfaces would be ensured in tokamaks if their toroidal symmetry were exact---including self-consistent plasma effects.   But, their confinement requires careful design for stellarators since stellarators cannot have an exact continuous spatial symmetry.  Nevertheless, the stellarator path to a fusion power plant is far more certain and faster than for a tokamak.  

The attractiveness of stellarators for fusion power plants follows from the dominance of the externally-produced magnetic field, which:
\begin{enumerate}
\item Provides robust passive stability.
\item  Allows reliable computational optimization.
\item Has an order of magnitude more degrees of freedom in the external magnetic field than in an axisymmetric field.
\end{enumerate}

Most fusion experiments are tokamaks; the largest is the (20-40 B\$) ITER, which was designed to produce net fusion power after 2035.  Tokamaks operate in a non-linear self-determined state, which requires active control.  Unfortunately, few control knobs are available.  Both diagnostics and controls become far more limited in burning plasmas than in existing tokamaks.  Loss of control results in disruptions and the transfer of the plasma current to relativistic-electron carriers.  Both can do major damage to the machine.  A solution is not known; an invention is required before tokamak power plants are possible.  The requirement of an invention makes estimates of time and certainty imponderable. 

Recognition of the problems of tokamaks with relativistic electrons (RE) and disruptions is increasing.   As noted in a 2019 review  \cite{Breizman:2019}: ``\emph{With ITER construction in progress, reliable means of RE mitigation are yet to be developed.}''   Machine damage from disruptions also appears more difficult to mitigate than previously thought.  In 2021, Nick Eidietis, who is a co-chair of the ITER-appointed Disruption Mitigation Task Force, reviewed the disruption situation in tokamaks \cite{Eidietis:2021}.  As noted in \cite{Boozer:steering}: ``\emph{Steering tokamak plasmas is commonly viewed as a way to avoid disruptions and runaway electrons. Plasma steering sounds as safe as driving to work but will be shown to more closely resemble driving at high speed through a dense fog on an icy road. The long time required to terminate an ITER discharge compared to time over which dangers can be foreseen is analogous to driving in a dense fog. The difficulty of regaining plasma control if it is lost resembles driving on an icy road.}"

Stellarators were thought to have a ``\emph{fatal flaw}" due to the absence of toroidal symmetry.  This can lead to a rapid  drift of the particles that form the fusion plasma across the magnetic field lines and unacceptably limit the energy confinement time---an even more fundamental problem than tokamak disruptions.  In 1981, Boozer \cite{Boozer:coord} developed a coordinate system and in 1984 a Hamiltonian description \cite{Boozer:H} of particle drifts in those coordinates.  These developments showed that a symmetry in the magnetic field strength confined particles as well as a symmetry in the vector $\vec{B}$ that represents all three components of the magnetic field.  In 1988, N\"uhrenberg and Zille \cite{N-Z} showed a stellarator can be designed so the magnetic field strength $B$ accurately approximates having a continuous symmetry even though the vector $\vec{B}$ cannot.  The ``\emph{fatal flaw}" of stellarators was eliminated.   

The most important result from the (1B\$) W7-X stellarator is that computational design works for stellarators even through a major change in configuration and scale \cite{Pedersen:2021}.  Tokamaks are designed by extrapolating from one generation of experiments to another.  The self-consistent non-linear state of tokamak plasmas gives no other option.

When time and certainty of success are critical, reliable computational design is vastly preferable to empirical extrapolations:
\begin{enumerate}

\item Experiments build in conservatism.

Even apparently minor changes in design are not possible and therefore remain unstudied. Major changes are risky even when going from one generation of experiments to another.

\item  Experiments are built and operated over long periods of time. 

Several decades are common.  A fast-paced program is inconsistent with many generations of experiments.

\item The cost of computational design is many orders of magnitude smaller than building a major experiment.

Innovative conceptual designs of stellarator power plants would cost $\approx$10M\$/year, ($\approx$2\% of U.S. fusion program). Much better designs appear possible.

4.  Extrapolations are dangerous when changing physics regimes.

Tokamak examples are (i) plasma control in ignited versus non-ignited plasmas and (ii) the formation of a current of relativistic electrons during a disruption.

\end{enumerate}

\begin{figure}
\centerline{ \includegraphics[width=3.0in]{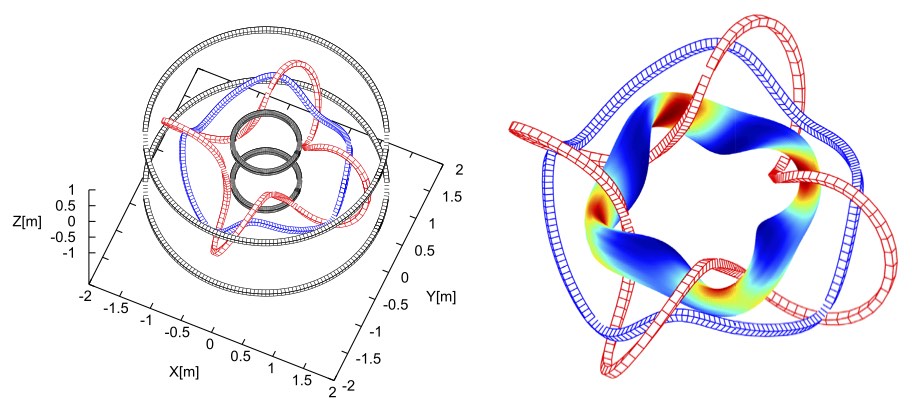} }
\caption{The large helical ripple in the magnetic field required in stellarators can be exploited to allow easy access to the plasma chamber and quick changes of internal components.  Although no one has optimized stellarator coils for open access, Yamaguchi's solution \protect\cite{Yamaguchi}  proves this is possible. Mathematics guarantees that all of the coils except the plasma-encircling red coil can be replaced by coils shaped like picture frames.  Picture-frame coils can be located in removable wall sections. }
\label{fig:Open-coil}
\end{figure}

The computational design of stellarators is not only desirable but also required:  
\begin{enumerate}

\item Well-confined magnetic field lines and particle drift trajectories are not automatic as in axisymmetric tokamaks.

\item Design optimization is subtle because of the large size of the optimization space.

This space is the fifty external magnetic-field distributions that can be produced with adequate efficiency, the same efficiency as the plasma shaping fields of tokamaks.  Efficiency mean the ratio of the magnetic field strength at the coils to that at the plasma.  A space of fifty degrees of freedom is too large to be fully explored, but an unlimited frontier invites discovery and invention.

\item  Designs can consider attractive plasma states that have no desirable tokamak analogue.

Fueling by pellet injection could be eased by having good confinement in only the outer third of the plasma. Transport could be controlled using internal transport barriers.

\item Unlike tokamaks, coils that allow open access to the plasma appear possible, Figure \ref{fig:Open-coil}.
\end{enumerate}

The efficiency of magnetic field distributions is limited because the coils must be located behind the blankets and shields that surround the plasma.  The blanket is where tritium is produced from lithium and the shields protect the superconducting coils from neutron damage.   The choice of magnetic field distributions that are controlled in a design is determined not only by their efficiency of production but also  by the sensitivity of plasma properties to them.  This sensitivity can differ by orders of magnitude and, especially in tokamaks, in sometimes surprising ways.  The issues of efficient field production and plasma sensitivity are discussed in \cite{Stellarator-design}.

As noted, tokamak plasmas require far more control, but the degrees of freedom to provide that control are far fewer.  The currents in the poloidal field coils of a tokamak, Figure \ref{fig:stellarators-tokamaks}, must depend on time.  A major control problem arises since the timescale for magnetic fields to penetrate through the blanket and shields, approximately half a second, can be far longer than the timescale for the plasma to evolve in undesirable situations.   In addition the natural decay time for the plasma current in ITER is a thousand seconds.  To shut down the plasma faster requires pulling magnetic flux out of the plasma using the transformer coils.  This can be much faster, but creates current profiles in the plasma that can cause disruptions.  Sixty seconds is thought to be the fastest disruption-free shutdown time for ITER.  The coil currents in stellarators need not evolve but can when a better design results.  Unlike magnetic fields produced by plasma currents, those produced by external coils do not need to be removed to shut the plasma down.

\begin{figure}
\centerline{ \includegraphics[width=2.8in]{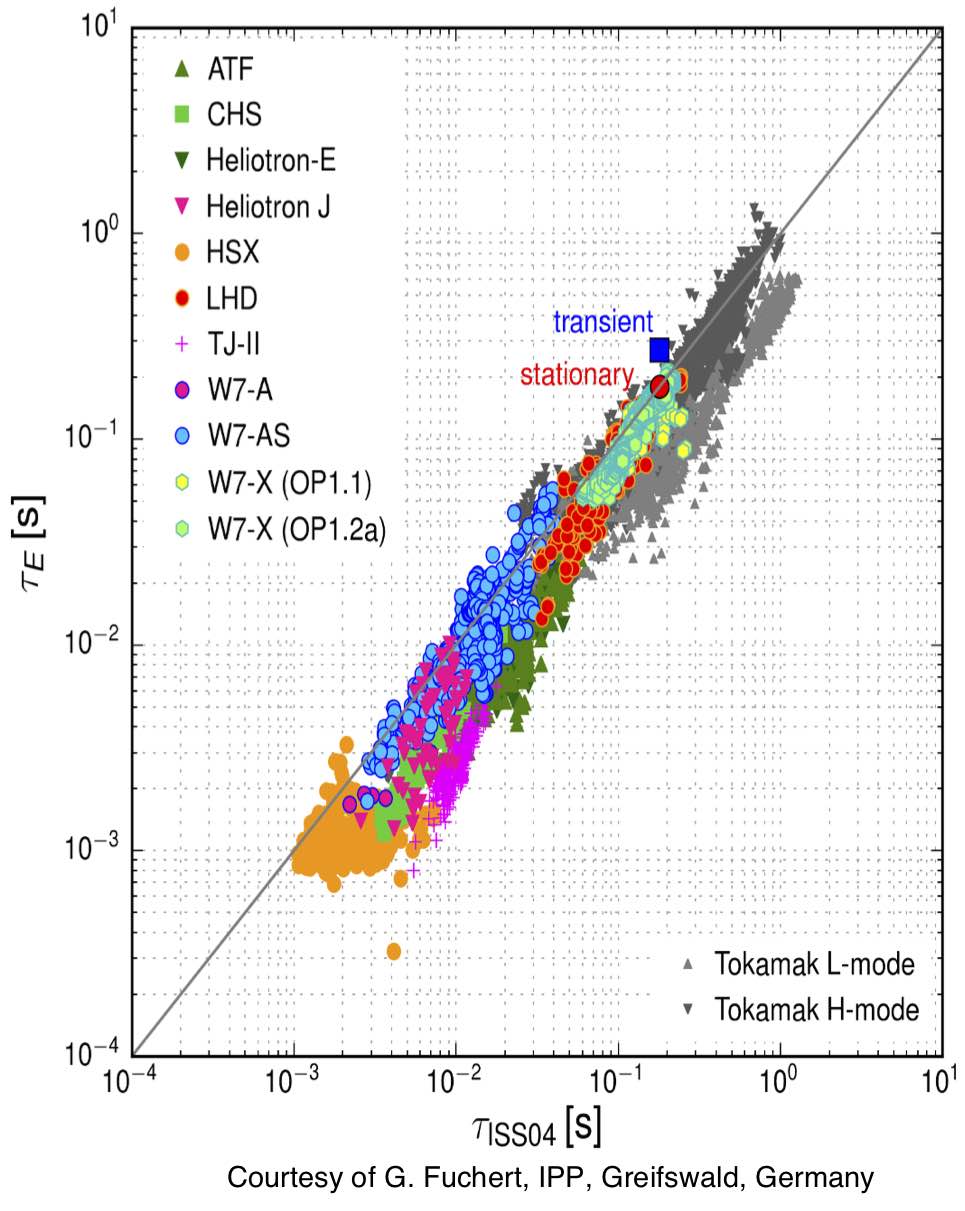} }
\caption{Empirically the tokamaks and stellarators have the same scaling of their energy confinement time.  Both obey what is called gyro-Bohm scaling \protect\cite{CO2-Stell}.}
\label{fig:transport}
\end{figure}

Empirical confinement times for energy in stellarators and tokamaks fits the same scaling law, Figure \ref{fig:transport}, which is given by gyro-Bohm diffusion within a dimensionless scaling factor \cite{CO2-Stell}. Extrapolations of the transport observed in long-pulse W7-X experiments yield attractive reactors.  

\begin{figure}
\centerline{ \includegraphics[width=3.0in]{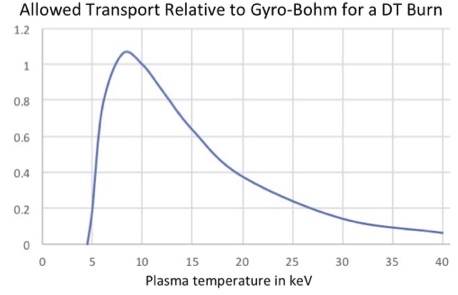} }
\caption{The empirical behavior of transport and the temperature dependence of the deuterium-tritium reactivity corrected for bremsstrahlung losses makes the required confinement of a self-sustaining fusion burn highly dependent on the plasma temperature, $T$.}
\label{Opt-T}
\end{figure}

The required energy confinement to maintain a DT burn in a fusion power plant depends on the plasma temperature, $T$.  Assuming the plasma transport is gyro-Bohm and ratio of the plasma to the magnetic field pressure is held fixed, the optimal temperature is $T\approx$10 keV, Figure \ref{Opt-T}.   Stellarator power plants could operate at $T\approx$10 keV, but current drive and the Greenwald limit on tokamak density force tokamaks power plants \cite{Stell-path} to have $T\approx$40 keV.  This and the limited plasma control makes obtaining adequate confinement much more difficult in tokamaks than in stellarator power plants.

Toroidal plasmas, whether in tokamaks or stellarators, need a system, called a divertor, which controls their contact with the surrounding chamber walls.  Divertors have requirements that appear contradictory.  They must concentrate the outflowing plasma that has reached the plasma edge into localized divertor chambers where pumps are located.  These pumps remove the helium ash and maintain a steady-state balance with the deuterium-tritium fueling. On the other hand, divertors cannot concentrate the outflowing heat into the divertor chambers because the average power density on the walls should be as high as technically possible to reduce the cost of fusion power.  The Watts of nuclear power striking a  square meter of the walls must be sufficient to pay for all the structures behind it.   

The solution to the contradictory demands on a divertor is detachment, which means the plasma flowing towards the divertor chambers radiates most of its energy content before it enters the chamber.  

Two types of divertors have been considered for stellarators: resonant and non-resonant.

\begin{figure}
\centerline{ \includegraphics[width=2.0in]{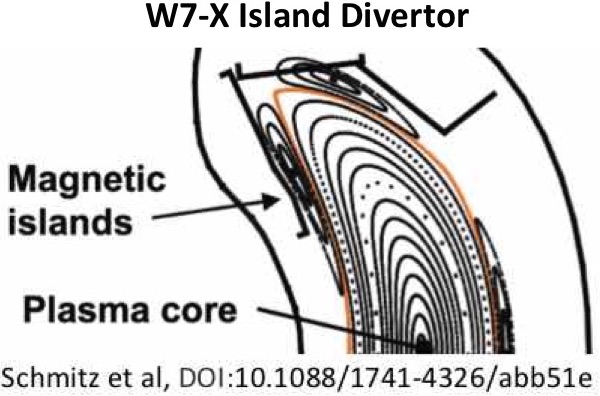} }
\caption{Resonant divertors require a specific twist of the magnetic field lines so an island in the magnetic field lines can be produced to define the divetor.  W7-X uses this type of divertor.}
\label{Res-div}
\end{figure}

Resonant divertors  utilize an island chain at plasma edge, Figure \ref{Res-div}.  They have been studied in W7-X and have demonstrated attractive longterm detachment properties \cite{Pedersen:2021}.  The achievement of robust steady-state detachment remains a major issue for tokamaks.

\begin{figure}
\centerline{ \includegraphics[width=2.5in]{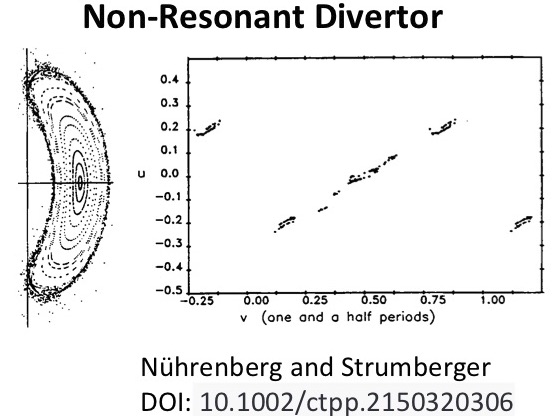} }
\caption{Stellarators tend to have an outermost confining magnetic surface.  Outside of that surface magnetic field lines tend to strike the walls in helical stripes \protect\cite{Strumberger:1992}, which can be used to define the location of the divertor chambers.  }
\label{Non-res-div}
\end{figure}

Non-resonant divertors \cite{Strumberger:1992} arise naturally in a stellarator, Figure \ref{Non-res-div}. There is an outermost magnetic surface that confines the plasma.  Outside that surface are generally Cantori which define tubes of magnetic flux that go from the plasma edge to the divertor chambers \cite{Punjabi}.  In numerical simulations, these flux tubes are observed to strike the same places on the wall (in helical stripes) even when important properties of the field are changed \cite{Bader}.

A major issue for both tokamaks and stellarators is the production of adequate tritium in the blanket \cite{Tritium:2021}.  As discussed in \cite{Stell-path}, stellarators have properties that better address a number of the tritium self-sufficiency issues:
\begin{enumerate}
\item Absence of disruption forces allows thinner structures and more tritium production.
\item  Open access coils allow fast changes is blanket structure to allow studies of multiple designs.
\item  The radial dependence of transport could be adjusted to make tritium use more efficient using shallow pellet injection.
\end{enumerate}


\section{Summary}

Major studies of the problem of carbon-dioxide increase, such as  \cite{IRENA:2021,WEO:2021}, focus on a strategy of ending the use of fossil fuels by 2050. This strategy ignores the millennia-long natural persistence of CO$_2$ once emissions end and the low cost of developing better options relative to the cost of deploying existing options.  

Emotional calls that focus on a fast elimination of fossil fuels to end climate change mislead the public. Without the deployment of a large scale system for direct air capture of carbon dioxide, the effects of CO$_2$ on the climate will be worse during the next several centuries than they are now. 

Fortunately, the need of direct air capture of CO$_2$ has been recognized \cite{DAC}, and perhaps efforts to solve the problem of an elevated CO$_2$ concentration will not be further impeded by the moral hazard of ``\emph{reducing humanity's will to cut emissions in the near term}" \cite{NAS-CO2}.

Arbitrarily large sums can be expended and time wasted on ill organized development programs.  The same can be said for ill conceived deployments.  Development and deployment can occur on a fast timescale when there is will and appropriate organization.  This is illustrated by the development of a Covid vaccine and the distribution to all American adults who wanted it in just over a year.  Decisions on which options should have expedited development and when to deploy the best existing options are not simple.  These decisions should be based reason rather than emotion.  Whatever the decisions may be, careful planning and organization are required in their implementation.

Nuclear energy is the carbon-free source that is neither intermittent nor localized in its places of application.  Fission energy could be deployed now on whatever scale is needed, and fission power plants could be made more suitable to the varied needs by further development.  Nevertheless, fusion energy has fundamental advantages in avoiding dangers such as the proliferation of nuclear weapons and long-lived radioactive wastes.

The fusion of deuterium and tritium is in principle the most attractive option for producing carbon-free energy.  Stellarators are far better poised than tokamaks for a fast and more certain development of a fusion power plant. 

The annual cost of an aggressive but well-organized minimal-time program to develop fusion energy would likely be less than ten billion dollars.  Typical designs for fusion power plants produce a gigawatt of electricity and should cost no more that ten billion dollars to be cost competitive with fission.  A ten-year construction period would cost a billion dollars a year.  The first of a kind machine may cost several times more, and several machines of different types should be built to mitigate risks. Research on material and construction concepts could be a billion dollars a year.

Ten billion dollars a year is a lot of money but much smaller than the thirty-eight billion dollars a year that Germany spends each year subsidizing green energy \cite{German-subsidy}  and tiny compared to four trillion dollars a year said to be needed to terminate the use of fossil fuels \cite{WEO:2021}.   An aggressive fusion program would have technological spinoffs just as did the eight-year Apollo program to land and return a person from the moon.  Obvious areas are better high-temperature superconductors, improved techniques for three-dimensional manufacture of large components, and better materials.

The first three to five years of a minimum-time stellarator program should be focused on computational conceptual-design, which would cost approximately 10M\$/year---a thousand times less than the annual cost of the construction period of a minimal-time fusion program.  Ten million dollars a year is only about 2\% of the present U.S. fusion program.

The absence of an aggressive program on the computational conceptual designs of stellarators anywhere in the world defies reason.  Perhaps the recognition of the importance of option development in Secretary Granholm's announcement on direct air capture of CO$_2$ \cite{DAC} will foster rational considerations on broader questions associated with the solution to the CO$_2$ problem.

\section*{Acknowledgements}
This work was supported by the U.S. Department of Energy, Office of Science, Office of Fusion Energy Sciences under Award Numbers DE-FG02-95ER54333, DE-FG02-03ER54696, DE-SC0018424, DE-SC0019479, and by grant 601958 within the Simons Foundation collaboration ``\emph{Hidden Symmetries and Fusion Energy}." 




\end{document}